\documentstyle[12pt,epsf]{article}
\begin{document}

\baselineskip 18pt

\newcommand{\sheptitle}
{Quadruple Yukawa Unification in the Minimal Supersymmetric Standard
Model}

\newcommand{\shepauthor}
{B. C. Allanach and S. F. King }

\newcommand{\shepaddress}
{Physics Department, University of Southampton\\Southampton, SO17 1BJ,
U.K.}

\newcommand{\shepabstract}
{Predictions for $m_t, \tan \beta, m_{\nu_\tau}$ are calculated for
quadruple third family $t-b-\tau-\nu_{\tau}$ Yukawa
unified models in the MSSM\@. The renormalisation group equations
for the 3 families of the MSSM, including the right handed neutrino,
are
presented.  For right handed tau neutrino Majorana masses that are
bigger than
$10^{11}$~GeV, the tau neutrino mass is consistent with present
cosmological bounds. The $m_t, \tan \beta$
predictions are approximately equivalent to those in triple third
family Yukawa unified models.}

\begin{titlepage}
\begin{flushright}
SHEP 95-16 \\
\end{flushright}
\vspace{.4in}
\begin{center}
{\large{\bf \sheptitle}}
\bigskip \\ \shepauthor \\ \mbox{} \\ {\it \shepaddress} \\
\vspace{.5in}
{\bf Abstract} \bigskip \end{center} \setcounter{page}{0}
\shepabstract
\end{titlepage}

The origin of the Yukawa couplings of the quarks and leptons
is one of the major puzzles facing the standard model.
In the context of supersymmetry (SUSY), and in particular the
minimal supersymmetric standard model (MSSM), some simplifications
may occur. The reason for this is simply that the low-energy gauge
couplings, when evolved using the renormalisation group equations
(RGE's) of the MSSM, converge at a scale $M_X=10^{16}-10^{17}$~GeV,
which hints at some further stage of unification at this scale.
Unified or partially unified gauge groups often constrain
not only the gauge couplings but also the Yukawa couplings of the
theory, and offer the possibility of understanding the low-energy
Yukawa couplings in terms of some simple pattern of Yukawa couplings
at the high-energy scale $M_X$.

It was realized some time ago that the simplest
grand unified theories (GUTs) based on SU(5)
predict the Yukawa couplings of the bottom quark
and the tau lepton to be equal at the GUT scale \cite{SU(5)},
\begin{equation}
\lambda_{b}(M_{X})=\lambda_{\tau}(M_{X})
\label{btau}
\end{equation}
where
$M_{X}\sim 10^{16}$~GeV.
Assuming the effective low energy theory below $M_{GUT}$ to be
that of the minimal supersymmetric standard model (MSSM)
the boundary condition in Eq.\ref{btau} leads to a physical
bottom to tau mass ratio $m_b/m_{\tau}$ that is in good agreement with
experiment
\cite{btauold}. Spurred on by recent LEP data which is
consistent with coupling constant unification,
the relation in Eq.\ref{btau} has recently been the subject of
intense scrutiny using increasingly sophisticated levels of
approximation \cite{btaunew,bandb}.

One may take this idea a stage further and consider theories based
on SO(10) or SU(4)$\otimes$SU(2)$_L\otimes$SU(2)$_R$
\cite{Yuk,patimass,otherpati}
which predict the Yukawa couplings of the top quark,
bottom quark and tau lepton all to be equal at the unification scale,
\begin{equation}
\lambda_{t}(M_{X})=\lambda_{b}(M_{X})=\lambda_{\tau}(M_{X}).
\label{tbtau}
\end{equation}
In such theories the top and bottom Yukawa couplings run almost
identically down to low energies, and the observed mass splitting
between the top and bottom is ascribed to a large ratio of
vacuum expectation values (VEVs) of the two Higgs doublets
of the MSSM, where the Higgs doublet which gains the large
VEV couples to the top quark, and the Higgs doublet which gains
the small VEV couples to the bottom quark. This ratio of VEVs
is conventionally defined to be $\tan \beta \equiv v_2/v_1$,
so that top-bottom-tau Yukawa unification predicts large
$\tan \beta \approx m_t/m_b$.

In this letter we shall include the tau neutrino Yukawa coupling,
which is predicted in some models to be equal to that of the
other third family Yukawa couplings
at the unification scale $M_X$:
\begin{eqnarray}
\lambda_t(M_X) = \lambda_b(M_X) = \lambda_\tau(M_X) =
\lambda_{\nu_\tau}(M_X)  & \equiv &
\lambda_{33}(M_X). \label{quadyukunified}
\end{eqnarray}
We refer to this as quadruple Yukawa unification.
As we shall see, the extra neutrino couplings can influence
the predictions which follow from top-bottom-tau Yukawa unification
if the right-handed neutrinos have a Majorana mass $M<M_X$.
Such right-handed neutrino Majorana masses may occur in the
two-loop Witten mechanism, for example \cite{Witten}.

The trilinear superpotential for the MSSM including a right handed
neutrino is
\begin{equation}
W = Y^U u Q H_2 + Y^D d Q H_1 + Y^E e L H_1 + Y^N \nu L H_2 +
\mbox{h.c},
\label{3famsuperpot}
\end{equation}
where $Y^U$, $Y^D$, $Y^E$ and $Y^N$ label the up quark, down quark,
charged lepton and neutrino Yukawa matrices
respectively\footnote{Family and gauge indices have been suppressed in
Eq.\protect\ref{3famsuperpot}.}. When the Higgs particles obtain their
vacuum expectation values (VEVs) $\langle H_1 \rangle = v \cos \beta$
and $\langle H_2 \rangle = v \sin \beta$, the neutrinos acquire
Dirac masses $m^D$.

The one loop coupling evolution of the Yukawa couplings in
Eq.\ref{3famsuperpot} was calculated in the $\overline{MS}$
renormalisation scheme, using an analysis of general
superpotentials performed by Martin and Vaughn \cite{MandV}:
\begin{eqnarray}
\frac{\partial g_i}{\partial t} &=& \frac{b_i g_i^3}{16 \pi^2}
\nonumber \\
\frac{\partial Y^U}{\partial t} &=& \frac{Y^U}{16 \pi^{2}} \left[
\mbox{Tr}
\left( 3 Y^U {Y^U}^{\dagger} + {Y^N}^\dagger Y^N \right) + 3
{Y^U}^{\dagger} Y^U +
{Y^D}^{\dagger} {Y^D}
- \left( \frac{13}{15} g_{1}^{2} + 3g_{2}^{2}
+ \frac{16}{3} g_{3}^{2} \right )\right] \nonumber \\
\frac{\partial {Y^D}}{\partial t} &=& \frac{{Y^D}}{16 \pi^{2}} \left[
\mbox{Tr}
\left( 3{Y^D} {Y^D}^{\dagger} + {Y^E} {Y^E}^{\dagger} \right) +
{Y^U}^{\dagger} Y^U + 3
{Y^D}^{\dagger} {Y^D} - \left( \frac{7}{15} g_{1}^{2} + 3
g_{2}^{2} + \frac{16}{3} g_{3}^{2} \right) \right] \nonumber \\
\frac{\partial {Y^E}}{\partial t} &=& \frac{{Y^E}}{16 \pi^{2}} \left[
\mbox{Tr}
\left( 3 {Y^D} {Y^D}^{\dagger} + {Y^E} {Y^E}^{\dagger} \right) + 3
{Y^E}^{\dagger} {Y^E} + {Y^N}^\dagger {Y^N} - \left( \frac{9}{5}
g_{1}^{2} + 3 g_{2}^{2} \right)
\right] \nonumber  \\
\frac{\partial Y^N}{\partial t} &=& \frac{Y^N}{16 \pi^2} \left[
\mbox{Tr} \left( 3 {Y^U}^\dagger Y^U + {Y^N}^\dagger Y^N \right)
 + 3 {Y^N}^\dagger Y^N + {Y^E}^\dagger Y^E - \left(
\frac{3}{5}g_1^2 + 3 g_2^2 \right) \right]
, \label{RG1nuR}
\end{eqnarray}
where $b_i=(33/5,1,-3)$, $t = \ln \mu$ and $\mu$ is the
$\overline{MS}$
renormalisation scale.

The see saw mechanism assumes the neutrino mass terms are of the
form
\begin{equation}
\left[ \overline{(\nu_L)}\ \overline{(\nu_R)^c} \right]
\left[\begin{array}{cc}
0 & m^D/2 \\
(m^D)^T/2 & M \\
\end{array}\right] \left[\begin{array}{c}
(\nu_L)^c \\
(\nu_R) \ \\
\end{array}\right] + \mbox{h.c.} =
m^D \overline{(\nu_L)} \ (\nu_R) + M \overline{(\nu_R)^c} \ (\nu_R)
+ \mbox{h.c.}
, \label{seesawmatrix}
\end{equation}
where $M >> m^D$ is the Majorana
mass of the right handed neutrino. In general, $m^D$ and $M$ are 3
by 3 matrices in family space.
In quadruple Yukawa unified models, the
Dirac masses $m^D$ of $\nu_e$
and $\nu_\mu$ are several orders of magnitude smaller than those of
the third family, similar to the charged fermions.
It is therefore a good approximation to consider the
third family alone and drop smaller Yukawa couplings.

Once the small couplings have been dropped, Eqs.\ref{RG1nuR} reduce to
the
RGEs derived in \cite{nubtaumass}:
\begin{eqnarray}
16 \pi^2 \frac{\partial g_i}{\partial t} &=& b_i g_i^3 \nonumber \\
16 \pi^{2} \frac{\partial \lambda_{t}}{\partial t} &=& \lambda_{t}
\left[ 6\lambda_{t}^{2} +  \lambda_{b}^{2}  + \theta_R
\lambda_{\nu_\tau}^2 -
\left( \frac{13}{15}
g_{1}^{2} + 3g_{2}^{2} + \frac{16}{3}g_{3}^{2} \right)
\right] \nonumber \\
16 \pi^{2} \frac{\partial \lambda_{b}}{\partial t} &=&
\lambda_{b} \left[ 6
\lambda_{b}^{2} + \lambda_{\tau}^{2} + \lambda_{t}^{2} -
\left(
\frac{7}{15} g_{1}^{2} + 3g_{2}^{2} +
\frac{16}{3} g_{3}^{2} \right) \right] \nonumber \\
16 \pi^{2} \frac{\partial \lambda_{\tau}}{\partial t} &=&
\lambda_{\tau}
\left[ 4 \lambda_{\tau}^{2} + 3 \lambda_{b}^{2} + \theta_R
\lambda_{\nu_\tau}^2
- \left( \frac{9}{5} g_{1}^{2}
+ 3g_{2}^{2} \right) \right] \nonumber \\
16 \pi^{2} \frac{\partial \lambda_{\nu_\tau}}{\partial t} &=&
\lambda_{\nu_\tau}
\left[ 4 \theta_R \lambda_{\nu_\tau}^{2} + 3 \lambda_{t}^{2} +
\lambda_{\tau}^2
- \left( \frac{3}{5} g_{1}^{2}
+ 3g_{2}^{2} \right) \right], \label{3rdfamnuReqs}
\end{eqnarray}
where $\theta_R \equiv \theta(t - \ln M)$ takes into account the
large mass suppression of the right-handed neutrino loops at scales
$\mu<M$. Thus we integrate out loops involving right-handed neutrinos
at $M$,
but retain the Dirac Yukawa coupling $\lambda_{\nu_{\tau}}$ which
describes the coupling of left to right-handed neutrinos.
The running procedure to determine the low energy masses is then to
run down the (Dirac)
neutrino Yukawa coupling $\lambda_{\nu_{\tau}}$ from
$M_X$ to low-energies using the above RGEs.
Then at low-energies we
use the usual
see-saw mechanism to determine the mass of the physical light
tau neutrino.

The procedure to extract the predictions from quadruple Yukawa
unification is as follows.
Values of $M$, $m_b(m_b)$, $\alpha_s(M_Z)$ and $M_X$ (the unification
scale)
were chosen as free parameters. Values of a scale $\Lambda$ and
$\lambda_{33} (M_X)$ are
taken. The gauge couplings $g_i$ are determined at $M_X$ by using the
input values
$\alpha_1(M_Z)^{-1}=58.89$,
$\alpha_2(M_Z)^{-1}=29.75$ and $\alpha_3(M_Z)=0.10-0.13$. The gauge
couplings were then run to $\Lambda$
by using the standard model RG equations
including 5 quark flavours and no scalar fields. The whole
superparticle
spectrum of the MSSM is assumed to lie at $\Lambda$ as an
approximation and the gauge
couplings are determined at $M_X$ by using the RGEs for the MSSM
\cite{bandb}. Although exact gauge unification is not imposed, it is
nevertheless approximately realised.
$\Lambda$ is taken to be $m_t \sim M_{susy}$  and
Eq.\ref{quadyukunified} is imposed
as a boundary condition\footnote{Note that in all of the predictions,
altering $M_{susy}$ to 1 TeV
makes only a negligible difference and so we are justified in taking
$M_{susy}=m_t$. }. The third family
and gauge
couplings are evolved from $M_X$ to $\Lambda$ using
Eq.s\ref{3rdfamnuReqs}.
\begin{figure}
\begin{center}
\leavevmode
\hbox{%
\epsfxsize=5.5in
\epsfysize=3.25in
\epsffile{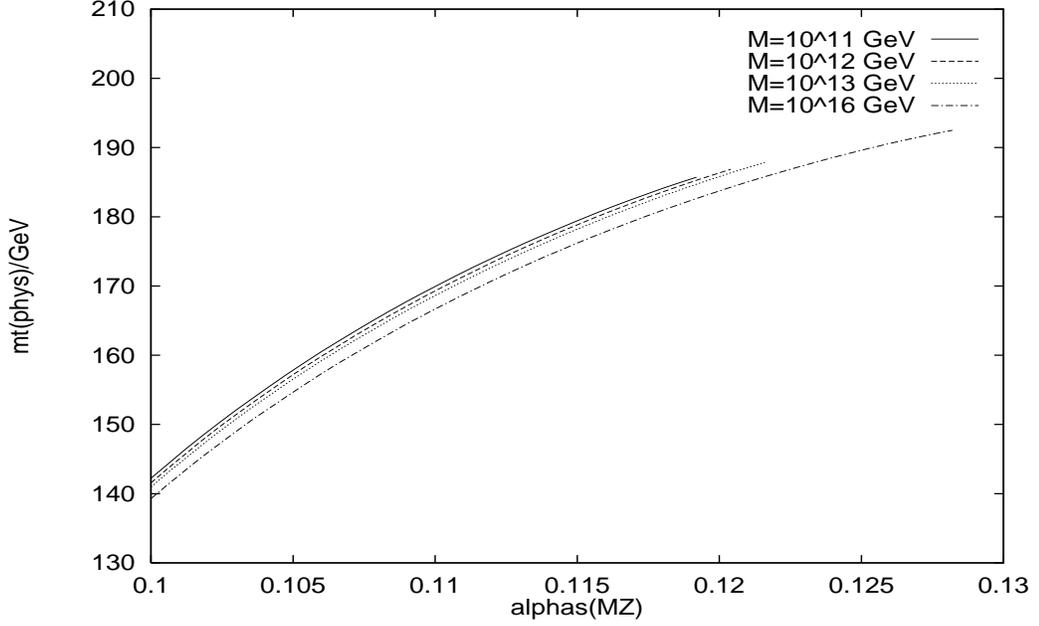}}
\end{center}
\caption{Physical $m_t$ predicted for various $M$, $M_X=10^{16}$~GeV
and $m_b=4.25$~GeV.}
\label{fig:mtopnuR}
\end{figure}

The various couplings at $\Lambda$ were determined
as follows. The running masses of the fermions
$m_{b,\tau}(\Lambda)$, were determined by running them
up from their mass shell values
$m_{f}(m_f)$ with effective 3~loop~QCD~$\otimes$~1~loop~QED
\cite{3loopqcdi,3loopQCDii,3loopQCDiii,guts}.
This enables us to calculate $\cos \beta$
at $\Lambda$ from the $\lambda_\tau(\Lambda)$ predicted by quadruple
Yukawa unification:
\begin{equation}
\cos \beta = \frac{\sqrt{2} m_\tau (\Lambda)}{v \lambda_\tau(\Lambda)}.
\end{equation}
The determination of $\cos \beta$ allows us to make the following mass
predictions from the Yukawa couplings calculated at the scale
$\Lambda$,
\begin{eqnarray}
m_t (\Lambda) &=& \frac{\sqrt{2} v \lambda_t (\Lambda)}{\sin \beta}
\nonumber \\
m_b (\Lambda) &=& \frac{\sqrt{2} v \lambda_b (\Lambda)}{\cos \beta}
\nonumber \\
m_{\nu_\tau} (\Lambda) &=& \frac{\sqrt{2} v \lambda_{\nu_\tau}
(\Lambda)}{\sin \beta}
\label{btpredictions}
\end{eqnarray}
Values of $\Lambda$ and $\lambda_{33}(M_X)$ are searched through until
$m_t=\Lambda$ is a prediction of Eq.\ref{btpredictions}, and
$m_b(\Lambda)$ is
predicted by Eq.\ref{btpredictions} to be the empirically derived
value
obtained by running $m_b (\mu)$ up from $m_b (m_b)$ as explained
above.
The light tau neutrino mass can now be predicted by diagonalising
Eq.\ref{seesawmatrix} in the one family case and extracting the small
mass eigenvalue:
\begin{equation}
m_{\nu_\tau} (m_{\nu_\tau}) = \frac{{m^D_{\nu_\tau}}(m_t)^2}{4 M}.
\label{physicalnutaumass}
\end{equation}
This equation is approximately correct because the Dirac mass does not
renormalise significantly from $m_t$ to $M_Z$ and below that scale,
effective QCD$\otimes$QED does not renormalise neutrino masses. We
thus
have a prediction for $m_t$ but whereas the $m_t$ referred to here is
the
running one,
it can be related as in \cite{bandb} to the physical mass by
\begin{equation}
m_t^{phys} = m_t \left( m_t \right) \left[ 1 + \frac{4}{3 \pi}
\alpha_3 \left( m_t \right) + O \left( \alpha_3^2 \right) \right].
\end{equation}

\begin{figure}
\begin{center}
\leavevmode
\hbox{%
\epsfxsize=5.5in
\epsfysize=3.25in
\epsffile{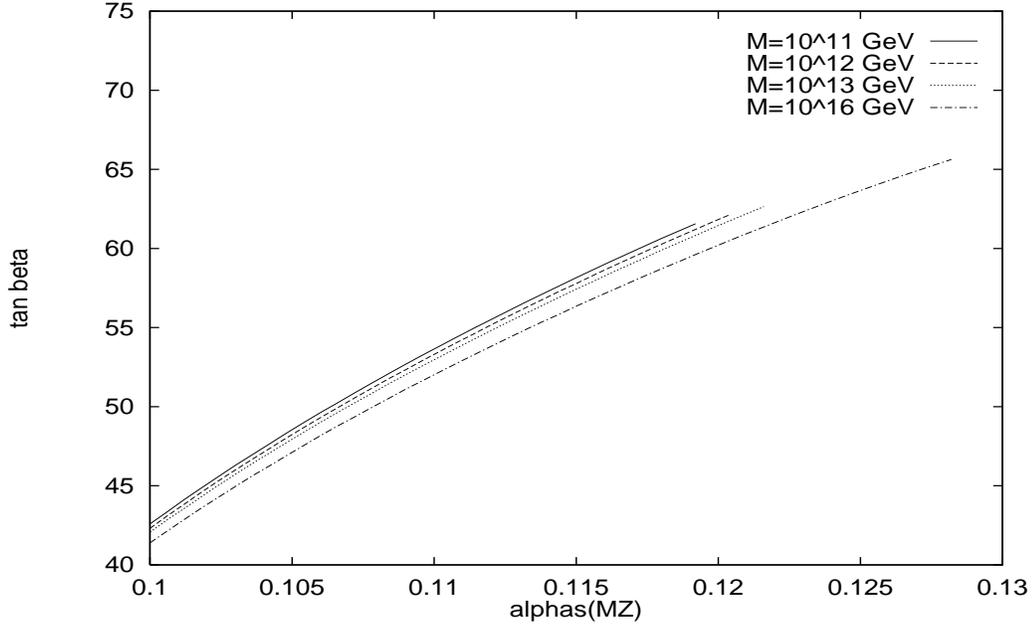}}
\end{center}
\caption{$\tan \beta$ predicted for various $M$
and $M_X=10^{16}$~GeV and $m_b=4.25$~GeV.}
\label{fig:tanbnuR}
\end{figure}
\begin{figure}
\begin{center}
\leavevmode
\hbox{%
\epsfxsize=5.5in
\epsfysize=3.25in
\epsffile{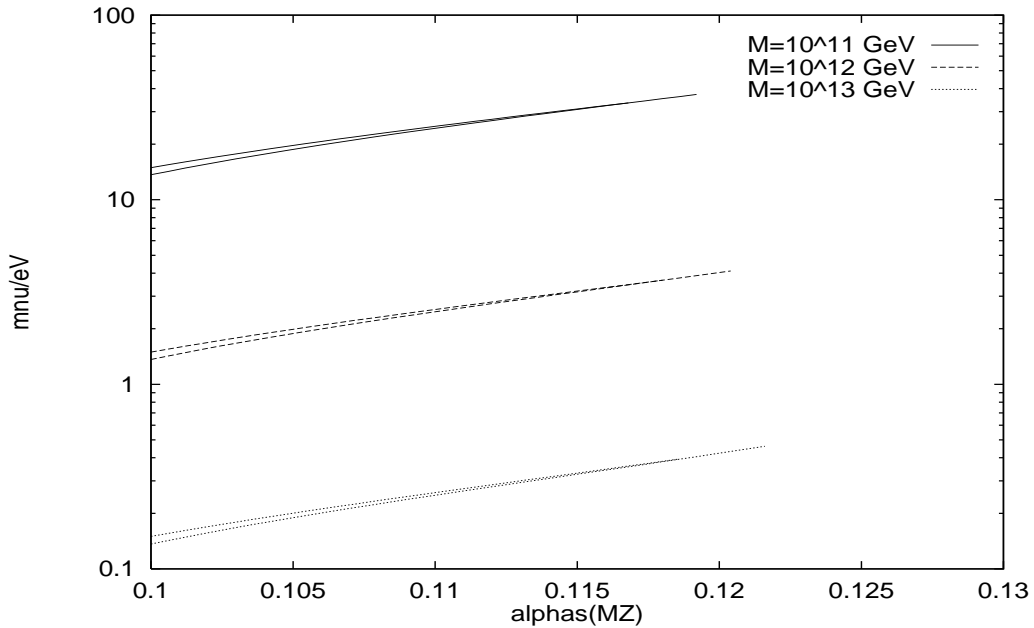}}
\end{center}
\caption{Physical tau neutrino mass predicted for
$m_b=4.25$~GeV and various $M, \alpha_S (M_Z)$. The upper lines
correspond to $M_X = 10^{16}$~GeV and
the lower lines to $M_X = 10^{17}$~GeV.}
\label{fig:mnuR}
\end{figure}
\begin{figure}
\begin{center}
\leavevmode
\hbox{%
\epsfxsize=5.5in
\epsfysize=3.25in
\epsffile{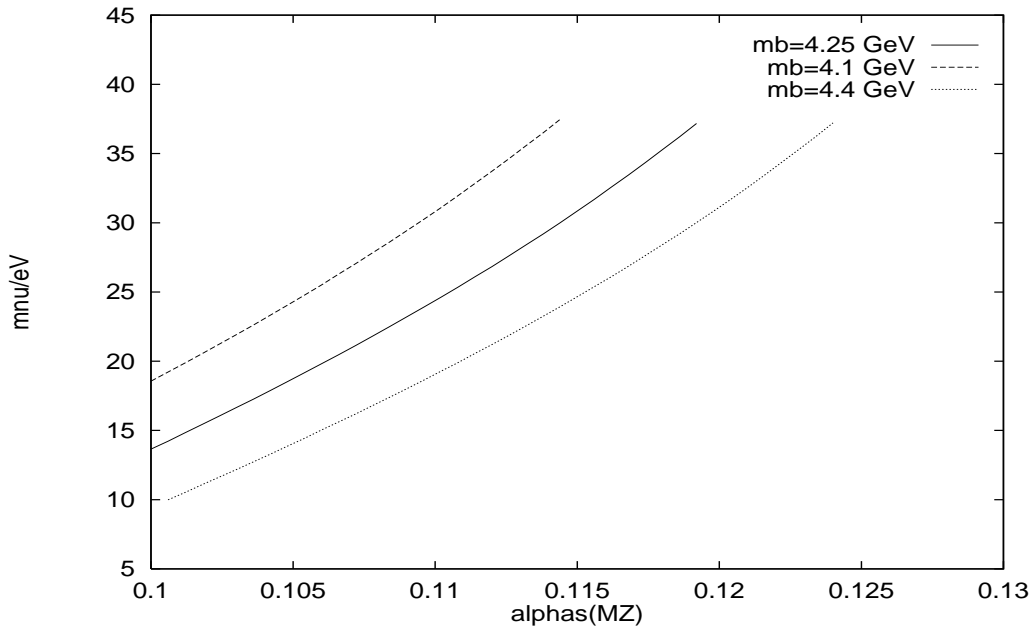}}
\end{center}
\caption{Physical tau neutrino mass predicted for  $M = 10^{11}$~GeV,
$M_X = 10^{16}$~GeV and various
$m_b, \alpha_S (M_Z)$.}
\label{fig:mnumbR}
\end{figure}

Fig.\ref{fig:mtopnuR} displays the difference in the $m_t$
predictions\footnote{Note that in Fig.s
\protect\ref{fig:mtopnuR}-\ref{fig:mnumbR}
the curves stop short for differing $\alpha_S (M_Z)$ because at these
values,
quadruple Yukawa unification is not possible with perturbative Yukawa
couplings (these are constrained to be less than 5.0.)}
for different $M$.
$M=10^{16}$~GeV corresponds to integrating out the right handed
neutrino at $M_X$ and so reduces to the previously
studied case of triple Yukawa unification in the MSSM
\cite{so10nonren,ourpaper}. As the figure shows, including the right
handed neutrino Yukawa coupling makes only a small
difference of up to 3~GeV to $m_t$. $m_t$ is insensitive to
whether the unification scale is $10^{16}$ or $10^{17}$~GeV.
Experimentally derived errors on $m_b, \alpha_S (M_Z)$
provide a much larger variation in $m_t$.
Fig.\ref{fig:tanbnuR} shows the difference in the $\tan \beta$
prediction when the right handed neutrino is included. Again, only a
small deviation from the triple Yukawa unification prediction is
observed. $m_b$ and $\alpha_S(M_Z)$ uncertainties provide a larger
deviation in the prediction \cite{patimass}. $\tan \beta$ is also
insensitive to
whether $M_X=10^{16}$ or $10^{17}$~GeV.
Because $\tan \beta$ and $m_t$ do not change significantly once the
right handed neutrinos has been taken into
account, it is reasonable to state that previous third family
calculations based on
triple Yukawa unification in the MSSM are valid in this scheme also.

Fig.\ref{fig:mnuR} shows the physical masses of the tau neutrino for
$M=10^{11,12,13}$~GeV. For $M=10^{10}$~GeV, $m_{\nu_\tau} > 150$
eV and so is excluded on cosmological grounds. In fact, the
cosmological bound
\begin{equation}
\sum_{i=e,\mu,\tau} m_{\nu_i} < 100 \mbox{~eV} \label{cosmobound}
\end{equation}
translates into a bound
on the Majorana mass: $M>10^{10}$~GeV. Note that $M>10^{13}$~GeV
implies that the tau neutrino would not be massive enough to observe
with present experiments. The choice of $M_X$ does not make an
appreciable difference to the tau neutrino mass.
The effect of the empirical range of $m_b=4.1-4.4$~GeV is shown in
Fig.\ref{fig:mnumbR} for $M=10^{11}$~GeV. $m_{\nu_\tau}$ can vary
by up 50 percent at low $\alpha_S (M_Z)$ for different values of
$m_b(m_b)$. Similar plots are obtained when $M_X$ is set equal to
$10^{17}$~GeV. Different $M$ values reproduce similar plots, with
the mass scaled as in Fig.\ref{fig:mnuR}.

The high value of $\tan \beta$ required for triple or quadruple Yukawa
unification is not stable under
radiative corrections unless some other mechanism such as extra
approximate symmetries are invoked.
$m_t$ may have been overestimated, since for high $\tan \beta$, the
equations for the running of the Yukawa couplings in the MSSM can get
corrections of a significant size from Higgsino--stop and
gluino--sbottom loops. The size of this effect depends upon the mass
spectrum and may be as much as 30~GeV \cite{so10nonren}.
 Not included in our analysis
are threshold effects, at low or high energies. These could alter our
results by several per cent and so it should be borne in mind that
all of the mass predictions have this uncertainty in them. It
is also unclear how reliable 3 loop perturbative QCD at 1~GeV is.

In conclusion, we have derived 3 family RGEs for the MSSM that include
the neutrino Yukawa coupling. We impose quadruple Yukawa unification
and make predictions for $m_t$, $\tan \beta$ and $m_{\nu_\tau}$. The
values of $m_t$
and $\tan \beta$ predicted in this scheme are approximately equivalent
to
results from triple Yukawa unification for values of $M$ that do not
violate
the cosmological bound on the neutrino masses ($M>10^{10}$~GeV).
A range of $M=10^{11}-10^{12}$~GeV predicts a tau neutrino mass that
does not violate the cosmological bound in Eq.\ref{cosmobound} and
that could possibly be observed in present day experiments.
Is such a value of $M$ reasonable theoretically?
A survey has recently been made of the application of the
Witten mechanism \cite{Witten} to various models \cite{stringyseesaw},
where it was seen that usually rather low values of $M$ are found.
Interestingly, out of several models examined,
supersymmetric SU(4)$\otimes$SU(2)$_L\otimes$SU(2)$_R$ gave the
highest $M$ generated by the Witten mechanism \cite{stringyseesaw}.
In that model, $M \sim M_X/(10^5-10^6)$ and so $M=10^{11}$~GeV
could result if $M_X \sim 10^{17}$~GeV. This model
\cite{patimass,otherpati} also predicts quadruple Yukawa unification
and so provides a predictive and simple scheme of viable tau neutrino
mass generation.

\end{document}